\documentclass[aps,prl,twocolumn,preprintnumbers,nofootinbib]{revtex4-2}
\usepackage[dvipsnames]{xcolor}
\definecolor{red}{rgb}{0.9, 0,0}
\definecolor{cerulean}{rgb}{0., 0.42,0.9}
\definecolor{navy}{rgb}{0.05, 0.05,0.8}

\usepackage[colorlinks]{hyperref}
\hypersetup{
colorlinks = true,
citecolor  = red,
linkcolor  = navy
}
\usepackage{mathtools}
\usepackage{amsmath}
\usepackage{amssymb}
\usepackage{amsfonts}
\usepackage{txfonts}
\usepackage{graphics}
\usepackage{graphicx}
\usepackage{endnotes}
\usepackage{slashed}

\newcommand{\Lpi}{\Lambda_{\pi}}
\newcommand{\mZp}{m_{Z'}}
\newcommand{\dUV}{d_{\mathrm{UV}}}
\newcommand{\gD}{g_{\chi}}
\newcommand{\gq}{g_{q}}

\newcommand{\mchi}{m_{\chi}}

\begin{document}
\title{A Warped Extra Dimensional Candidate for the LZ 248 keV Event}
\author{Vincent S. H. Lee$^{*}$}

\affiliation{Department of Physics, University of California, Berkeley, Berkeley, CA 94720, USA}
\affiliation{Department of Physics, University of California, San Diego, La Jolla, CA 92093-0319, USA}
\author{Lisa Randall$^{\dagger}$}
\affiliation{Gravity, Spacetime, and Particle Physics (GRASP) Initiative, Harvard University, 17 Oxford Street, Cambridge, MA 02138, USA.}
\preprint{N3AS-26-020}
\begin{abstract}
	The recent result by the LUX-ZEPLIN (LZ) experiment of a single nuclear-recoil event at $248\,\mathrm{keV}$ suggests an interpretation in terms of inelastic scattering between dark matter (DM) and the xenon target. The widely considered thermal Higgsino DM with $m_{\chi}\sim$~TeV and mass splitting $\delta \sim 350$~keV is essentially excluded as it would be captured efficiently by the Sun, with a resulting neutrino flux from its annihilation in excess of the IceCube limit, and expected recoils above the LZ nominal search window, where none were observed. However we argue that although these tensions apply to a Higgsino with constrained cross section, they do not necessarily to the more widely possible set of inelastic candidates that alternative solutions to the hierarchy problem might suggest. We show that a new massive gauge boson, $Z'$~\cite{DiMauro:2026ldr} with a candidate DM sector, with  Majorana mass splitting of order $\sim 10^{-7}$ of the Dirac mass, evades current constraints.  We propose an inelastic dark matter (iDM) candidate that can arise naturally in a Randall-Sundrum warped extra dimension model. The DM candidate involves a vector-like fermion confined to  the TeV brane and a bulk field. The brane Lagrangian contains Dirac mass terms that are also naturally exponentially suppressed. The left-handed component of the bulk field has a Majorana mass with origin on the UV brane, while the interaction with Standard Model nucleons is mediated by a massive dark Kaluza–Klein (KK) gauge boson whose mass is also automatically in the right range. We propose a concrete model of thermal DM with $m_{\chi}\sim$~TeV in an RS geometry with warp factor $\pi kR\simeq 31.7$ and the $Z'$ a KK mode with natural TeV-ish scale mass in which the ratio of neutrino to quark coupling is also natural, giving rise to a mass splitting of $\delta \sim 300$~keV, and show how it could explain the LZ signal while being consistent with existing experimental constraints.
\end{abstract}

\maketitle

\begingroup\renewcommand{\thefootnote}{\fnsymbol{footnote}}
\footnotetext[1]{\href{mailto:vincentszehimlee@berkeley.edu}{vincentszehimlee@berkeley.edu}\\ \llap{$\dagger$\hspace{3.3pt}}\href{mailto:randall@g.harvard.edu}{randall@g.harvard.edu}}
\endgroup

\textbf{Introduction.}---The LUX-ZEPLIN (LZ) collaboration has recently reported a single nuclear-recoil event at $248\,\mathrm{keV}$ with a global significance of $2.6\sigma$ and a maximal local significance of $3.4\sigma$~\cite{LZ:2026axp}. If this is due to dark matter (DM), the absence of lower-energy nuclear-recoil events hints that the scattering process could be inelastic in nature. In such a model, DM includes two nearly degenerate mass states and the lighter one upscatters to the heavier in the collision, so that only halo particles above a velocity threshold can scatter at all~\cite{Tucker-Smith:2001myb, Bramante:2016rdh}. This is known as inelastic dark matter (iDM). Writing the two states as $\chi_1$ and $\chi_2$ with masses $m_1<m_2$, $m_\chi\simeq m_1\simeq m_2$ and $\delta\equiv m_2-m_1$, the event points to $\delta\simeq100$--$350\,\mathrm{keV}$ for $m_\chi=0.4$--$4\,\mathrm{TeV}$~\cite{LZ:2026axp}.

Within days of the release, several groups identified the natural iDM candidate as a Higgsino~\cite{Freese:2026sga, Fan:2026kxx, Wu:2026nhi, DiMauro:2026ldr, Smirnov:2026aqk, Du:2026guj} (see Refs.~\cite{Su:2026rwz, Lou:2026idn, DiMauro:2026ldr, Unwin:2026rdp, Nomura:2026qyq, Visinelli:2026kgt} for other possibilities). Its neutral-current coupling is off-diagonal between the two Majorana mass eigenstates, so tree-level $Z$ exchange produces only the inelastic transition, with a cross section fixed by the gauge coupling at $G_F^2\mu_n^2/2\pi\simeq7.4\times10^{-39}\,\mathrm{cm}^2$~\cite{Rodd:2026tyn}. The event then requires $\delta\simeq370$--$420\,\mathrm{keV}$ for the standard halo model and $\delta\simeq470$--$490\,\mathrm{keV}$ if a high-velocity component from the Large Magellanic Cloud is included~\cite{Rodd:2026tyn}.

Ref.~\cite{Pospelov:2026ewn} has, however, pointed out that this interpretation appears to be in tension with solar observations. The Sun accelerates infalling DM by its gravitational pull, so a $Z$-coupled doublet could be captured efficiently, and the neutrinos from annihilation in the solar core via $\chi\chi\to W^+W^-$ and $\chi\chi\to ZZ$ could exceed the IceCube limit~\cite{IceCube:2025fcu} unless $\delta\gtrsim500$--$570\,\mathrm{keV}$,  in conflict with the Higgsino interpretation of the  LZ event. Moreover, Ref.~\cite{Rodd:2026tyn} separately demonstrated a possible tension between the Higgsino interpretation and the absence of events in the LZ high-energy sideband, whose acceptance is not public.

These tensions arise because of the tight connection between the Higgsino and Standard Model (SM) parameters. Whereas the cross section for the SM finds an optimal mass splitting  at about 350 keV, a more general iDM candidate has more freedom to fit both the cross section and mass splitting more independently.  With a smaller mass splitting, the above constraints can be handily avoided.

Given that the DM being detected is at the weak scale, it seems reasonable that other models addressing the hierarchy problem might also have a suitable candidate. In light of the above constraints, the preferred DM candidate would be a SM singlet that scatters off nucleons through a dark mediator, so that annihilation into SM gauge bosons is suppressed~\cite{Arkani-Hamed:2008hhe, Batell:2009vb, Bramante:2016rdh}.  An example of such a model with a new massive $Z'$ gauge boson was investigated in detail in Ref.~\cite{DiMauro:2026ldr}. However, the minimal model still leaves open the questions of why $\delta$ should be small.   For the Higgsino, $\delta\simeq m_Z^2/M_{\mathrm{gaugino}}$ is naturally small owing to the heaviness of the gauginos~\cite{Rodd:2026tyn}, although even in that case an inordinately high Majorana mass for the gauginos was required~\cite{Fan:2026kxx}. A good DM candidate in alternative theories should also explain such a small parameter. Though technically natural as it can be explained by a DM number symmetry, an underlying explanation would be far more compelling.  

In this Letter, we argue that a warped extra dimension--in particular one addressing the hierarchy problem--can provide a natural origin for such a hierarchy. In a Randall-Sundrum (RS) geometry~\cite{Randall:1999ee}, a mass parameter on the UV brane is redshifted to the TeV scale on the IR brane, and a further exponential suppression arises whenever the transmission between the branes proceeds through a bulk field whose wavefunction is localized away from one of them. A Majorana splitting far below the Dirac mass is therefore generic for DM on the IR brane when the source of Majorana mass sits on the UV brane, with $\delta/\mchi\sim10^{-7}$ corresponding to order-one bulk mass parameters rather than to a small coupling. For instance, this mechanism was introduced for iDM by Ref.~\cite{Cui:2009xq} as potential explanations for the now-defunct  DAMA modulation signal~\cite{DAMA:2008jlt}, and admits several realizations differing in how the breaking is communicated to the brane. Below we provide one explicit model and show that it can naturally explain the LZ signal, while satisfying existing experimental constraints.

This Letter is organized as follows. We first present the model and specify its parameters: a thermal SM singlet on the IR brane of an RS geometry whose Majorana splitting is generated on the UV brane, and whose interactions with nucleons are mediated by a heavy $U(1)_x$ gauge boson. This fermion is charged in a ``Higgsless" model~\cite{Csaki:2003dt, Csaki:2003zu} 
with the UV boundary condition breaking the gauge symmetry so the lightest gauge mediator is a Kaluza–Klein (KK) mode, whose warp factor suppressed mass is also set by the (roughly weak scale) IR scale. %
We then show that the model reproduces the LZ event and the observed relic abundance, while satisfying  solar, collider, lifetime, and sideband constraints. Finally we conclude.

\begin{figure}[t]
	\centering
	\includegraphics[width=\linewidth]{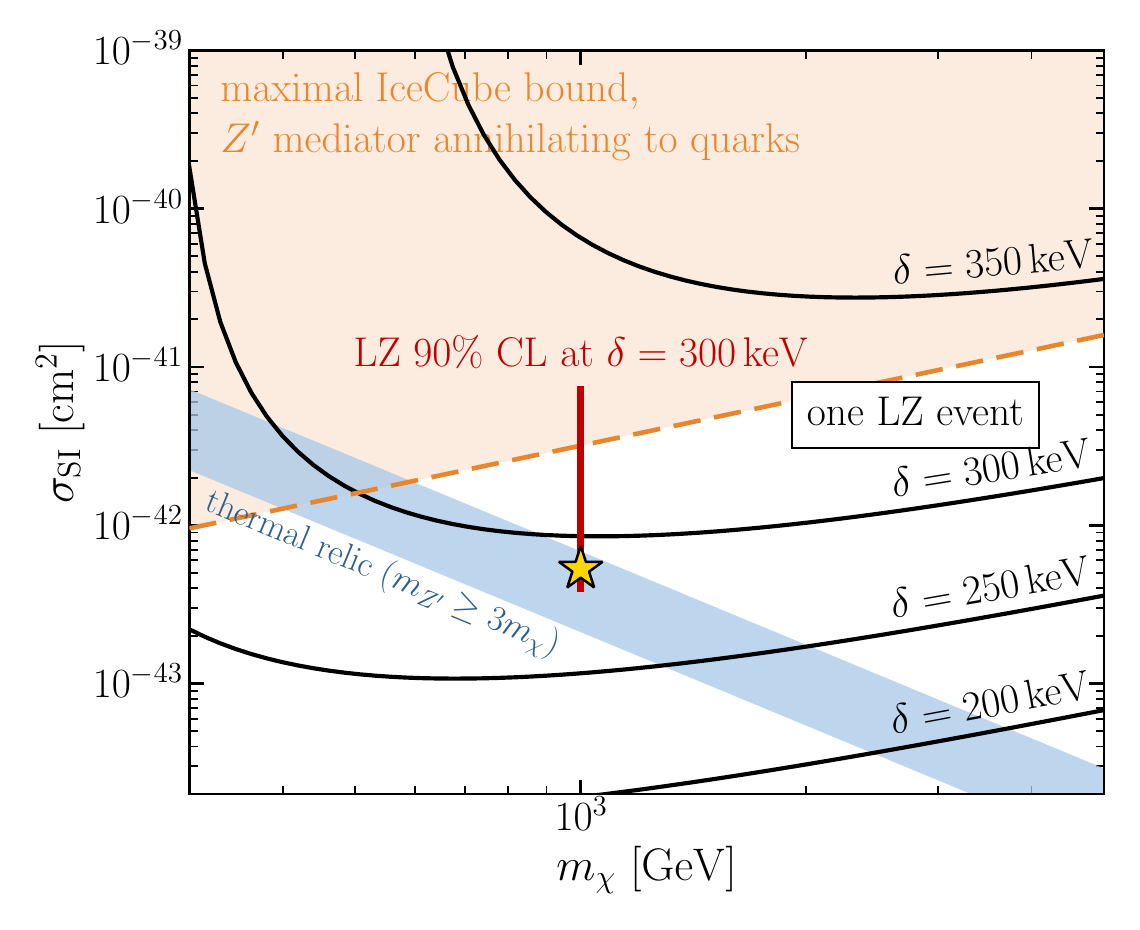}
    \caption{The DM--nucleon scattering cross section as a function of the DM mass. The blue band is the cross section of a thermal relic annihilating through the $Z'$, Eq.~\eqref{eqn:thermal_sigma}, for $\mZp\ge3\mchi$. The black curves give one event in the LZ search region for $\delta=200$, $250$, $300$, and $350\,\mathrm{keV}$. The red bar is the two-sided $90\%$ confidence interval of LZ at $\mchi=1\,\mathrm{TeV}$ and $\delta=300\,\mathrm{keV}$~\cite{LZ:2026axp}. The orange region is the IceCube limit on neutrinos from solar capture, for a $Z'$ mediator with isoscalar couplings annihilating to quarks, rescaled from Ref.~\cite{Pospelov:2026ewn} as described below. It is a maximal bound: the annihilation rate is set to its saturation value $C/2$, which requires capture and annihilation to have reached equilibrium. Away from equilibrium the rate is smaller and the excluded region recedes upward. The star marks the benchmark of this Letter.}
	\label{fig:thermal_plane}
\end{figure}

\textbf{Warped inelastic dark matter.}---We consider the following model. The geometry is a slice of $\mathrm{AdS}_5$ on the orbifold $S^1/Z_2$~\cite{Randall:1999ee}, with metric $ds^2=e^{-2k|y|}\eta_{\mu\nu}dx^\mu dx^\nu-dy^2$ for $y\in[-\pi R,\pi R]$, where $k$ is the curvature, and $R$ the radius of the fifth dimension. The UV brane sits at $y=0$ and the IR brane, on which the SM resides, at $y=\pi R$. The warp factor redshifts the (reduced) Planck scale, defined as $M_{\mathrm{Pl}}\equiv (8\pi G)^{-1/2}$, to $\Lpi\equiv M_{\mathrm{Pl}}e^{-\pi kR}$ on the IR brane, and we take $\pi kR=31.7$ and $k/M_{\mathrm{Pl}}=0.05$, so that $\Lpi=42\,\mathrm{TeV}$ and the first KK graviton, at $3.83\,ke^{-\pi kR}=8.0\,\mathrm{TeV}$, lies above its collider bound~\cite{ATLAS:2021uiz}. The warped-down curvature is given by $k\,e^{-\pi kR}=2.1\,\mathrm{TeV}$. This value of $\pi kR$ is in accordance with the desired masses of the $U(1)_x$ gauge boson $Z'$, which we introduce below, at $\sim5\,\mathrm{TeV}$, as well as the dark matter particle. The DM is a vector-like Dirac fermion $\chi=(\chi_L,\chi_R)^T$ on the IR brane with Dirac mass $\mchi$. We take $\mchi=1\,\mathrm{TeV}$, the mass at which LZ report their two-sided interval on the cross section~\cite{LZ:2026axp}. It is also the mass singled out by the event itself since, as noted in Ref.~\cite{DiMauro:2026ldr}, a vector mediator with thermal relic abundance fixes the DM--nucleon scattering cross section to $6.5\times10^{-43}\,\mathrm{cm}^2\,(1\,\mathrm{TeV}/\mchi)^2$ in the heavy-mediator limit, which is the same as the scattering rate in LZ that matches the one observed event near $1\,\mathrm{TeV}$, as we show in Fig.~\ref{fig:thermal_plane}. %

The DM is assumed to be a SM singlet with vector-like charge $x_\chi$ under a $U(1)_x$ gauge symmetry. There is also a bulk fermion $S=(S_L,S_R)^T$ with the same $U(1)_x$ charge and bulk mass $ck$ for some $c\sim \mathcal{O}(1)$, and $(+,+)$ boundary conditions for $S_L$, with a Majorana mass for $S_L$ on the UV brane, where $U(1)_x$ is broken. The singlet action is written as~\cite{Cui:2009xq}
\begin{align}\label{eqn:singlet_action}
	S &= \int d^4x\,dy\,\sqrt{g}\Big\{-\tfrac{1}{4}Z'_{MN}Z'^{MN} + \bar S\,i\gamma^MD_MS + ck\,\epsilon(y)\,\bar SS \nonumber \\
	&\quad - \left(\tfrac{1}{2}\dUV\,\bar S^c_LS_L + \mathrm{h.c.}\right)\delta(y) \nonumber \\
	&\quad + \left[\bar\chi\,i\gamma^\mu D_\mu\chi - \mchi\bar\chi\chi - \left(\mu\,\bar\chi_RS_L + \mathrm{h.c.}\right)\right]\delta(y-\pi R)\Big\} \, ,
\end{align}
where $Z'_{MN}$ is the field strength of the bulk $U(1)_x$ gauge field $Z'_M$, $D_M=\partial_M-ig_5x_\chi Z'_M$ acts on $\chi$ and $S$ with the same charge, $\epsilon(y)$ is the sign function, $\dUV$ is the dimensionless coefficient of the UV-brane Majorana mass of $S_L$, and $\mu$, which sets the IR-brane mixing between $S_L$ and $\chi_R$, has mass dimension $1/2$, with $\mu^2$ taken to be of order $k$. In addition, $S_L$ is even and $S_R$ odd under the orbifold, and%
, writing $Z'_M=(Z'_\mu, Z'_5)$, $Z'_\mu$ obeys Dirichlet boundary conditions on the UV brane and Neumann on the IR brane~\cite{Csaki:2003dt}, so that the $U(1)_x$ is absent on the UV brane, allowing for a  Majorana mass for $S_L$. The lightest KK mode of the $Z'$ has mass set by the geometry, $\mZp=2.4\,ke^{-\pi kR}=5\,\mathrm{TeV}$~\cite{Csaki:2003dt}. This is the warped bulk fermion model in Ref.~\cite{Cui:2009xq}, with the electroweak doublet and the $Z$ replaced by a SM singlet charged under the $U(1)$ and a $U(1)_x$ KK gauge boson but in which no additional Higgs field or global DM $U(1)$ symmetry is required. (As an alternative, one could explicitly gauge the theory with a vev-dependent gauge boson mass).  We work in the mass basis of $S$, whose wave function is obtained by imposing the UV-brane Majorana mass as a boundary condition on the bulk profiles~\cite{Agashe:2015izu}. The brane term $\mu$ mixes $\chi_R$ with the KK modes of $S$, which are paired into Dirac fermions by their KK masses, so that the mass eigenstates are linear combinations of $\chi$ and $S$. Computing the splitting of $\chi$ to leading order in this mixing we find
\begin{equation}\label{eqn:warped_splitting}
	\delta = \frac{\mu^2\,e^{-2\pi kR}}{\dUV}\,e^{-2(c-1/2)\pi kR} \, ,
\end{equation}
in agreement with Ref.~\cite{Cui:2009xq}. We have verified  the accuracy of this approximation by diagonalizing the full $\chi$--$S$ mass matrix numerically and find that, for $\mchi$ below the KK scale, the exact splitting agrees with Eq.~\eqref{eqn:warped_splitting} to within $\mathcal{O}(10\%)$, a difference  readily absorbed by a small shift in $c$.

\begin{figure}[t]
	\centering
	\includegraphics[width=\linewidth]{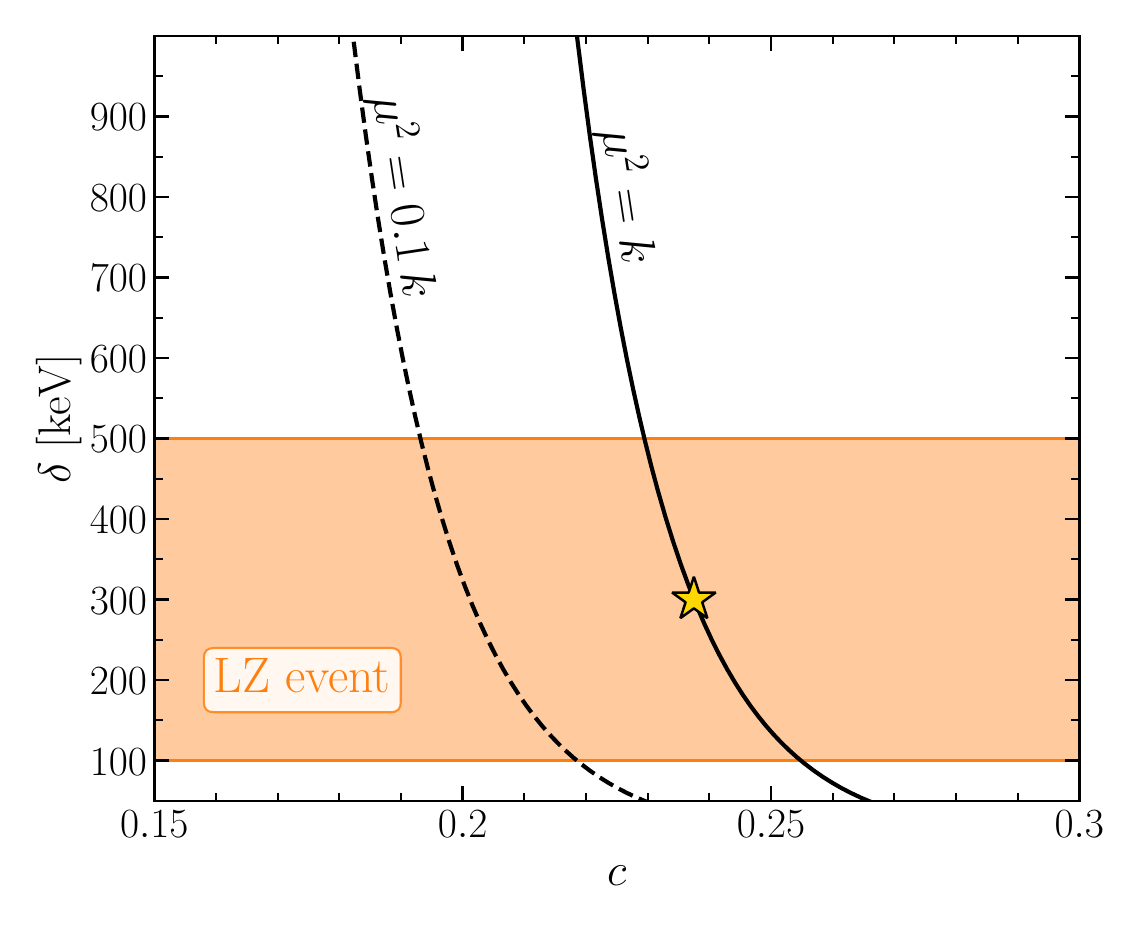}
	\caption{The mass splitting $\delta$ of the two Majorana components of $\chi$ as a function of the messenger bulk mass parameter $c$, from Eq.~\eqref{eqn:warped_splitting} with $\pi kR=31.7$, $\dUV=2$, and $\mu^2=k$ or $0.1\,k$. The shaded band marks the splittings that can account for the LZ event at $\mchi=1\,\mathrm{TeV}$: below $\sim 100\,\mathrm{keV}$ the predicted recoils fall mostly below the observed one~\cite{LZ:2026axp}, and above $\sim 500\,\mathrm{keV}$ no halo particle is fast enough to produce it, taking the velocity distribution including the Large Magellanic Cloud~\cite{Smith-Orlik:2023kyl}. The star is the benchmark of this Letter.}
	\label{fig:delta_vs_c}
\end{figure}

We thus see that an iDM model with a  small mass splitting is naturally generated with no small parameters in the action put by hand. %
We plot $\delta$ as a function of $c$ in Fig.~\ref{fig:delta_vs_c}. We take $\delta=300\,\mathrm{keV}$ as our benchmark: large enough that the local significance of the LZ fit remains close to its maximum~\cite{LZ:2026axp}, while small enough that the recoil spectrum does not populate the high-energy sideband~\cite{Rodd:2026tyn}, as we show below. This mass splitting is naturally obtained by setting $\dUV=2$, $\mu^2=k$ and $c=0.2375$ in Eq.~\eqref{eqn:warped_splitting}.

The $U(1)_x$ gauge boson $Z'$ has mass $\mZp$ and couples to the DM and to the quarks
\begin{equation}\label{eqn:zprime_couplings}
	\mathcal{L}\supset -Z'_\mu\left(\gD\,\bar\chi\gamma^\mu\chi + \gq\sum_q\bar q\gamma^\mu q\right) \, ,
\end{equation}
where $\gD$ and $\gq$ are the couplings to the DM and to each quark flavor. Only the product $\gD\gq$ enters the LZ rate and the relic abundance, and it is fixed by the latter below. We will discuss the bounds on individual couplings below.
Since we take $S$ to be charged, we assume spectators to address $U(1)$ anomalies. Moreover since the quarks are confined on the IR brane  we assume any mixed anomalies are cancelled by Chern-Simons terms, something readily accomplished in such a higher-dimensional model.

This model also generates a mixing between the $Z'$ and the $U(1)_Y$ gauge boson via quark loops, giving rise to the kinetic-mixing operator $-\tfrac{\epsilon}{2}Z'_{\mu\nu}B^{\mu\nu}$, where $Z'_{\mu\nu}$ and $B_{\mu\nu}$ are the $U(1)_x$ and $U(1)_Y$ field strengths, with~\cite{Holdom:1985ag, Cheung:2009qd, Gherghetta:2019coi}
\begin{equation}\label{eqn:epsilon}
    \epsilon=\frac{g'\gq}{16\pi^2}\sum_qY_q\ln\frac{\Lambda^2}{\mZp^2} \, ,
\end{equation}
the sum running over the quark chiralities and colors, with $g'=e/\cos\theta_W\simeq0.36$ the hypercharge gauge coupling and $Y_q$ the quark hypercharges, where we take the tree-level mixing to vanish at the cutoff scale $\Lambda\sim (24\pi^3)^{1/3}M_5e^{-\pi kR}$ of the theory~\cite{Chacko:1999hg} warped to the IR brane, where $M_5$ is the five-dimensional Planck mass given by $M_5^3=kM_{\mathrm{Pl}}^2$~\cite{Randall:1999ee}. For $\gq=0.3$ this gives $\epsilon\simeq0.03$. At $\mZp=5\,\mathrm{TeV}$ its effects on electroweak observables are suppressed by $m_Z^2/\mZp^2$ and negligible, but it gives the $Z'$ a coupling to every SM fermion, including $g_\nu=\epsilon g'/2\simeq5\times10^{-3}$ to neutrinos.

\textbf{The LZ event.}---We now examine how the LZ signal arises in this model. The $Z'$ mediates a spin-independent interaction between $\chi$ and nucleons, since its vector coupling to quarks sums coherently over the nucleus, and because its coupling to $\chi$ is off-diagonal the scattering is $\chi_1N\to\chi_2N$. For $\mZp\gg q$ the cross section per nucleon is~\cite{Cui:2009xq}
\begin{equation}\label{eqn:zprime_cross_section}
	\sigma_{\mathrm{SI}} = \frac{\mu_n^2}{\pi}\,\frac{(3\gD\gq)^2}{\mZp^4} \, ,
\end{equation}
where $\mu_n$ is the reduced mass between the DM and the nucleon. 

We now see the real freedom of the model, since the parameters are not fixed but can be chosen in accordance with the requisite LZ cross section for appropriate mass splitting and with thermal relic abundance as a supplementary constraint.  Since the $Z'$ couples universally to quarks, its couplings to protons and neutrons are equal and the interaction is isoscalar. For our benchmark, we set the product $\gD\gq=0.56$, such that Eq.~\eqref{eqn:zprime_cross_section} gives $\sigma_{\mathrm{SI}}=4.9\times10^{-43}\,\mathrm{cm}^2$. On the experimental side, LZ have interpreted the event as isoscalar spin-independent inelastic scattering and published a two-sided $90\%$ confidence interval on $\sigma_{\mathrm{SI}}$ at $\mchi=1\,\mathrm{TeV}$ as a function of $\delta$~\cite{LZ:2026axp}; at $\delta=300\,\mathrm{keV}$ it is the red bar in Fig.~\ref{fig:thermal_plane}, and our benchmark lies within it.

To compare with the LZ data at other masses, we compute the expected number of events. The inelastic recoil spectrum is given by~\cite{Rodd:2026tyn}
\begin{equation}\label{eqn:spectrum}
    \frac{dR}{dE_R}\propto F^2(E_R)\,\eta\big(v_{\min}(E_R)\big) \, ,
\end{equation}
where $E_R$ is the nuclear recoil energy, $F^2(E_R)$ is the Helm form factor with the parameters of Ref.~\cite{1996APh.....6...87L}, and $\eta(v_{\min})=\int_{v>v_{\min}}d^3v\,f(\mathbf v)/v$ is the mean inverse speed of halo particles above the minimum velocity $v_{\min}$ needed to produce the recoil, for the velocity distribution $f$ in the detector frame, and $v_{\min}$ given by~\cite{Tucker-Smith:2001myb}
\begin{equation}\label{eqn:vmin}
    v_{\min}(E_R)=\frac{1}{\sqrt{2m_AE_R}}\left(\frac{m_AE_R}{\mu_A}+\delta\right) \, ,
\end{equation}
with $\mu_A$ the DM--nucleus reduced mass. We take the standard halo model with the parameters of Ref.~\cite{Rodd:2026tyn} and the LZ nuclear-recoil efficiency~\cite{LZ:2026axp}. Integrating the spectrum over the LZ search region, $5.4$ to $270\,\mathrm{keV}$, for their exposure of $2.84$ tonne-years and setting the result to one event gives, for each $\delta$, the cross section shown by the black curves in Fig.~\ref{fig:thermal_plane}; at $\mchi=1\,\mathrm{TeV}$ they pass through the LZ intervals.
Here we see why this model can evade the Higgsino model constraints. The additional freedom in parameters ($\mZp$, $\gD$, $\gq$) allows $\delta$ to be independently set by the kinematics, with residual parameters fit to the cross section. However the model is not unconstrained. We next consider the additional restriction on parameters from requiring the correct thermal abundance.

\textbf{Relic abundance.}---The freeze-out relic abundance of $\chi$ is set by its annihilation rate into the SM plasma. Since $\mchi<\mZp$, annihilation into $Z'$ pairs is closed, and the abundance is instead set by the coannihilation $\chi_1\chi_2\to Z'^*\to q\bar q$ through the off-diagonal coupling. Its cross section is that of a Dirac fermion annihilating through an $s$-channel vector~\cite{Foguel:2024lca}
\begin{equation}\label{eqn:relic}
    \sigma_{12}(s) = \frac{3\,\gD^2\gq^2}{2\pi}\,\frac{s+2\mchi^2}{(s-\mZp^2)^2+\mZp^2\Gamma_{Z'}^2}\,\frac{1}{\sqrt{1-4\mchi^2/s}} \, ,
\end{equation}
where $s$ is the center-of-mass energy squared, the prefactor includes the sum over six quark flavors and three colors, and $\Gamma_{Z'}$ is the total width of the $Z'$,
\begin{equation}\label{eqn:zprime_width}
    \Gamma_{Z'} = \frac{\gD^2\mZp}{12\pi}\left(1+\frac{2\mchi^2}{\mZp^2}\right)\sqrt{1-\frac{4\mchi^2}{\mZp^2}} + \frac{3\,\gq^2\mZp}{2\pi} \, ,
\end{equation}
the first term from $Z'\to\chi\bar\chi$ and the second from $Z'\to q\bar q$. 

At freeze-out the DM is nonrelativistic, so $\sigma_{12}v$ is evaluated in the limit $s\to4\mchi^2$, where it becomes velocity independent. Since the freeze-out temperature of a weak-scale relic, $\mchi/20$ to $\mchi/30$, is far above $\delta$, $\chi_1$ and $\chi_2$ are equally populated at freeze-out, and because only $\chi_1\chi_2$ pairs annihilate, the cross section that enters the freeze-out calculation is $\sigma v_{\mathrm{eff}}=\tfrac12\sigma_{12}v$.
\begin{equation}\label{eqn:sigmav_eff}
    \sigma v_{\mathrm{eff}} = \frac{9\,\gD^2\gq^2\,\mchi^2}{\pi\left[(\mZp^2-4\mchi^2)^2+\mZp^2\Gamma_{Z'}^2\right]} \, .
\end{equation}
The relic abundance is related to the cross section by~\cite{Gondolo:1990dk}
\begin{equation}\label{eqn:relic_density}
    \Omega\, h^2 \simeq \frac{8.5\times10^{-11}\,\mathrm{GeV}^{-2}\,x_f}{\sqrt{g_*}\,\sigma v_{\mathrm{eff}}} \, ,
\end{equation}
where $x_f=\mchi/T_f$ with $T_f$ the freeze-out temperature and $g_*$ the number of relativistic degrees of freedom at freeze-out. Requiring the observed $\Omega h^2=0.12$ with $x_f=26.7$ and $g_*=94$ gives $\sigma v_{\mathrm{eff}}=2.3\times10^{-26}\,\mathrm{cm}^3/\mathrm{s}$, which for $\mZp=5\,\mathrm{TeV}$ fixes $\gD\gq=0.56$. 

Since both Eq.~\eqref{eqn:zprime_cross_section} and Eq.~\eqref{eqn:sigmav_eff} depend on the couplings only through $\gD\gq$, the relic abundance fixes the scattering cross section up to the distance of the $Z'$ from the annihilation pole,
\begin{equation}\label{eqn:thermal_sigma}
    \sigma_{\mathrm{SI}} = \sigma v_{\mathrm{eff}}\,\frac{\mu_n^2}{\mchi^2}\left(1-\frac{4\mchi^2}{\mZp^2}\right)^2 \, ,
\end{equation}
where $\sigma v_{\mathrm{eff}}\simeq2\times10^{-26}\,\mathrm{cm}^3/\mathrm{s}$ is nearly independent of $\mchi$. Our treatment of freeze-out is valid only when the $Z'$ mass is sufficiently far from the annihilation pole.  When the cross section varies rapidly with energy, the threshold value of Eq.~\eqref{eqn:sigmav_eff} no longer represents the thermal average, and the relic abundance can be resonantly enhanced~\cite{Gondolo:1990dk, Griest:1990kh}. Imposing $\mZp\ge3\mchi$, a thermal relic occupies the blue band in Fig.~\ref{fig:thermal_plane}. %

\textbf{Constraints.}---We now discuss the experimental constraints. As mentioned in the introduction, IceCube constrains DM captured in the Sun by searching for the neutrinos produced when it annihilates in the solar core. Ref.~\cite{Pospelov:2026ewn} recast this, for the Higgsino, as an upper limit on the annihilation rate into $W^+W^-$,
\begin{equation}\label{eqn:icecube_ww}
    \Gamma_{\mathrm{ann}}(\chi\chi\to W^+W^-) < 1.5\times10^{19}\,\mathrm{s}^{-1}
\end{equation}
at $\mchi\simeq1\,\mathrm{TeV}$.

 Our model couples DM to $W$ or $Z$ bosons only through the WZW-like Chern-Simons term, which is loop suppressed and thus gives at best a naive branching-ratio estimate of order $(g^2/16\pi^2)^2\simeq7\times10^{-6}$, where $g\simeq0.65$ is the $SU(2)_L$ gauge coupling, up to anomaly coefficients and kinematic and multiplicity factors.
DM primarily annihilates to quarks and, through the kinetic mixing, to neutrinos. We recast the estimate assuming the same capture calculation, but with different final states. IceCube derived the cross-section limit for every annihilation channel from the limit on the annihilation rate in that channel, so at fixed mass the ratio of the rate limits for two channels should correspond to  the ratio of their published cross-section limits. Reading these off at $1\,\mathrm{TeV}$~\cite{IceCube:2025fcu}, the $b\bar b$ limit is $20$ times the $W^+W^-$ limit and the $\nu\bar\nu$ limit is $0.02$ times it, giving approximate upper bounds $\Gamma_{\mathrm{ann}}(\chi\chi\to q\bar q)<3\times10^{20}\,\mathrm{s}^{-1}$ and $\Gamma_{\mathrm{ann}}(\chi\chi\to\nu\bar\nu)<3\times10^{17}\,\mathrm{s}^{-1}$.

The annihilation rate in the Sun is bounded by the capture rate, given by $\Gamma_{\mathrm{ann}}=\tfrac12C\tanh^2(t/\tau_{\mathrm{eq}})\le C/2$ at all times, where equality is only achieved once capture and annihilation have equilibrated~\cite{Pospelov:2026ewn}. We obtain the capture rate by rescaling the Higgsino calculation of Ref.~\cite{Pospelov:2026ewn}: the capture rate is proportional to the cross section and, for an isoscalar interaction, to $A^2$ rather than the Higgsino's $(A-Z)^2$, and the inelastic threshold restricts it to iron and heavier nuclei in both cases. Their capture rate of $4\times 10^{23}\,\mathrm{s}^{-1}$ at $\sigma=7.4\times10^{-39}\,\mathrm{cm}^2$ then becomes $C\simeq 4\times 10^{23}\times(4.9\times10^{-43}/7.4\times10^{-39})\times(A/(A-Z))^2_{\mathrm{Fe}}\simeq 9\times10^{19}\,\mathrm{s}^{-1}$ for our benchmark. Each annihilation removes two particles, so $\Gamma_{\mathrm{ann}}\le C/2\simeq 4.6\times10^{19}\,\mathrm{s}^{-1}$ whether or not capture and annihilation have equilibrated~\cite{Pospelov:2026ewn}. The quark channel, with branching fraction $\simeq1$, then gives $\Gamma_{\mathrm{ann}}(\chi\chi\to q\bar q)\le 4.6\times10^{19}\,\mathrm{s}^{-1}$, below the IceCube bound by a factor of $\sim 6$. Since the capture rate is proportional to the cross section and to the number density of DM, this is a bound on the cross section, $\sigma_{\mathrm{SI}}\lesssim 3.2\times10^{-42}\,\mathrm{cm}^2\,(\mchi/1\,\mathrm{TeV})$, which is the orange region in Fig.~\ref{fig:thermal_plane}. The neutrino channel has branching fraction $(g_\nu/\gq)^2/12$, so the IceCube $\nu\bar\nu$ bound gives $g_\nu<0.28\,\gq$, which is above $g_\nu=\epsilon g'/2\simeq5\times10^{-3}$ from kinetic mixing at the benchmark. Finally, anomalous decays to $W^+W^-$ are subdominant. The bound on the annihilation rate through
this channel is 20 times tighter than on quarks, but the branching
fraction is smaller by around five orders of magnitude as stated above. We note here that these bounds will be relaxed if equilibrium has not been attained. Additionally, since the current bound derived above is only cleared by a factor of 6, this indicates that DM-induced neutrinos from the Sun could be observed in the near future, but a careful analysis on whether equilibrium is achieved will be required.

In addition, at the LHC the $Z'$ would be produced from quarks and searched for as a dijet resonance or as missing momentum recoiling against a jet or photon. ATLAS exclude a $Z'$ coupling to quarks with universal $\gq$ only below $4.6\,\mathrm{TeV}$, even at $\gq=0.5$, the largest coupling in their scan~\cite{ATLAS:2019fgd}, so there is no dijet constraint at $\mZp=5\,\mathrm{TeV}$. Even if we take a slightly lower mass in their range the bound on $g_q$ would be very weak, of order $0.4$. The induced lepton coupling $\epsilon g'Y_\ell\simeq10^{-2}$ is far below the reach of dilepton searches at this mass~\cite{ATLAS:2026nqc}.

The $Z'$ contribution to the muon anomalous magnetic moment is $|\Delta a_\mu|\sim g_\mu^2m_\mu^2/(12\pi^2\mZp^2)$ with $g_\mu=\epsilon g'Y_\mu\simeq10^{-2}$~\cite{Queiroz:2014zfa}, six orders of magnitude below the current sensitivity of $6\times10^{-10}$~\cite{Aliberti:2025beg, Muong-2:2025xyk}.

The scattering event we describe is an upscattering from the lighter state $\chi_1$ to the heavier state $\chi_2$, which requires that the DM today be entirely $\chi_1$%
, i.e. that $\chi_2$ has decayed within the age of the universe. Since $\delta<2m_e$ and our coupling is purely vector-like, the only available SM final state is a neutrino pair, %
and the process is $\chi_2\to\chi_1\nu\bar\nu$. The decay rate is $\Gamma(\chi_2\to\chi_1\nu\bar\nu)\simeq\gD^2g_\nu^2\delta^5/40\pi^3\mZp^4$~\cite{March-Russell:2009vla}, which corresponds to a lifetime of $0.08\,\mathrm{Gyr}$ for our benchmark, so no $\chi_2$ survives today. %

We note that indirect constraints for models with such a small lepton coupling are readily satisfied, and that additional constraints on the RS sector are alleviated if the SM resides entirely on the brane as we have assumed.

Finally, we consider the issues with sideband events as identified in Ref.~\cite{Rodd:2026tyn}. LZ report no events in the $350$ to $590\,\mathrm{keV}$ sideband above their search window, which can be in tension with the interpretation of DM scattering for some choices of $m_{\chi}$ and $\delta$ if the predicted number of scattering events in this band for such choices is too large. The ratio of the expected number of events in the sideband, $N_{\mathrm{SB}}$, to that in the search region, $N_{\mathrm{SR}}$, is
\begin{equation}\label{eqn:sideband}
    \frac{N_{\mathrm{SB}}}{N_{\mathrm{SR}}} = \frac{\int_{350\,\mathrm{keV}}^{590\,\mathrm{keV}} dE_R\,\dfrac{dR}{dE_R}}{\int_{5.4\,\mathrm{keV}}^{270\,\mathrm{keV}} dE_R\,\dfrac{dR}{dE_R}} \, ,
\end{equation}
with $dR/dE_R$ from Eq.~\eqref{eqn:spectrum}, assuming equal acceptance in the two regions. For our benchmark we find $N_{\mathrm{SB}}/N_{\mathrm{SR}}\sim 0.08$, so no tension arises. Ref.~\cite{Rodd:2026tyn} finds that the Higgsino, whose larger $\delta\simeq377\,\mathrm{keV}$ places the peak of the spectrum inside the sideband, has this ratio of order several, in tension with the empty sideband. This does not happen in our model since a smaller $\delta$ is allowed.

This does however suggest an interesting test of the model when the full data set with three times the data is released. For this $\delta$, we expect 2 or 3 events in the full data set in the fiducial region and 0 or 1 event in the sideband. However, with more data the best fit mass splitting $\delta$ and cross section should be refit in accordance with the results. The current model can also be tested by searches for an 8 TeV KK graviton resonance.

\textbf{Conclusion.}---We have presented a model of inelastic dark matter in a warped extra dimension that accounts for the LZ $248\,\mathrm{keV}$ event. The DM is a SM singlet on the IR brane, its Majorana splitting $\delta/\mchi\sim10^{-7}$ is generated on the UV brane from an order-one bulk mass parameter, and its scattering is mediated by the lightest KK mode of a bulk $U(1)_x$ gauge boson. With $\mchi=1\,\mathrm{TeV}$, $\delta=300\,\mathrm{keV}$, and $\mZp=5\,\mathrm{TeV}$ the model reproduces the LZ rate and the relic abundance and evades all existing constraints.

\textbf{Acknowledgments.}---We thank Carlos Arg\"uelles-Delgado, Jaqueline Lodman, Rashmish K. Mishra,  Matthew B. Reece, Huy Tran, and Taewook Youn for helpful discussions.  V.L. is supported by the Network for Neutrinos, Nuclear Astrophysics and Symmetries (N3AS) through the National Science Foundation Physics Frontier Center, Grant No. PHY-2020275. L.R. is supported by the Gravity, Spacetime, and Particle Physics (GRASP) Initiative from Harvard University.  %

\bibliography{bibliography}
\end{document}